# Can  gravity  distinguish between Dirac and Majorana Neutrinos?

## S. A Alavi, A. Abbasnezhad

**Department of Physics, Hakim sabzevari  university,**

**P. O. BOX 397, Sabzevar, Iran.**

s.alavi@hsu.ac.ir , alaviag@gmail.com

The interaction of neutrinos with gravitational fields in the weak field regime at one loop to the leading order has been studied by Menon and Thalappilil. They deduced some theoretical differences between the Majorana and Dirac neutrinos. Then they proved that, in spite of these theoretical differences, as far as experiments are concerned, they would be virtually indistinguishable. We study the interaction of neutrinos with weak gravitational fields to the second order (at two loops). We show that there appear new neutrino gravitational form factors which were absent in the first-order calculations, so from a theoretical point of view there are more differences between the two kinds of neutrinos than in the first order, but we show that likewise they are indistinguishable experimentally.

**Key Words: Dirac and Majorana neutrinos, gravitational fields.**

## I- Introduction

The whole of  the  last  century  of  physics  is  recognized  by two main theories: Quantum Mechanics and   Relativity. The   physical   phenomena   in which gravitational   and quantum effects   appear simultaneously are very interesting both from theoretical  and  experimental point  of  view [1-13]. There  has  been  an extensive  of  research  in  theoretical  physics which brought out an unexpected interplay between general relativity and  quantum field theory. On the other hand many attempts have been  made  to  see  whether there are some   novel experimental or observational ways of studying quantized fields coupled to curved spacetime.  Interaction  of  quantum  particles  with  gravitational fields  is  one   the  interesting subjects at the interface of quantum mechanics and general relativity. Neutrino is one of the most mysterious and interesting particles in the universe. Neutrino physics is one of the most important     fields of research in high energy physics,  astrophysics and cosmology, see [14,15] for a recent review. Interaction of neutrinos with gravitational fields  and  the  distinguishability between   Dirac and Majorana neutrinos  are  very exciting issues in neutrino physics. The graviton-neutrino vertex to the first order (1-loop) has been studied in [16] using general symmetry principles. They  tried  to  understand  how  the  Majorana    and   Dirac neutrinos  could  be  different  as  far  as gravitational interaction is concerned. They found that in spite of theoretical differences the Majorana and Dirac neutrinos cannot be distinguished by gravitational interaction as far as experiments are involved. It is worth to mention that neutrino form factors have also been studied in [17,18]. In this work we study the graviton-neutrino vertex and gravitational neutrino transition form factors to the second order i.e., at 2-loops. We show that from theoretical perspective the second order calculations reveal more differences between Dirac and Majorana neutrinos than the first order but we also show that they are indistinguishable as far as experiments are considered. It is worth mentioning that it is shown in [19] that spin-gravity interaction can distinguish between Dirac and Majorana neutrino wave packets propagating in a Lense-Thirring background but it is pointed out in [20] that the treatment of Majorana neutrino in [19] is not valid so the claim stated in [19] does not follow, see also [21]. In Ref. [22-29], one can find some papers on the possibilities of distinguishing Dirac from Majorana neutrinos but not in the gravitational field.

## II. 2 loops calculations

For Dirac neutrinos, there are eighteen 2-loops graviton-neutrino vertex which are shown in Fig. (1) to Fig. (18) :

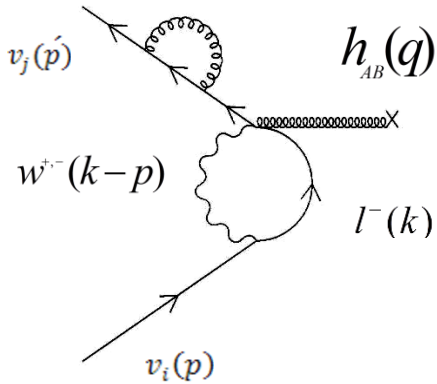

FIG. 1

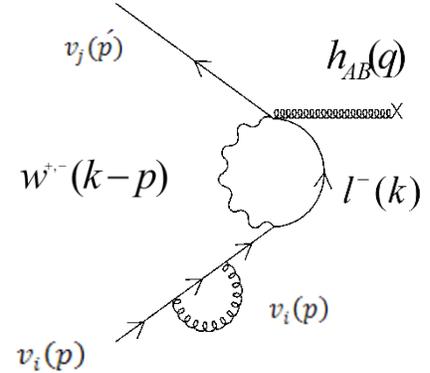

FIG. 2

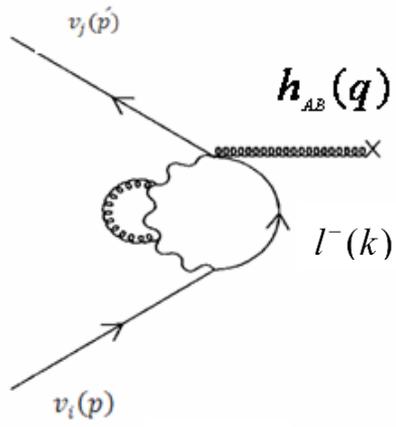

FIG. 3

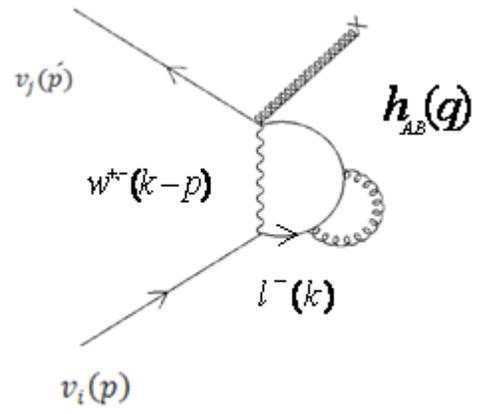

FIG. 4

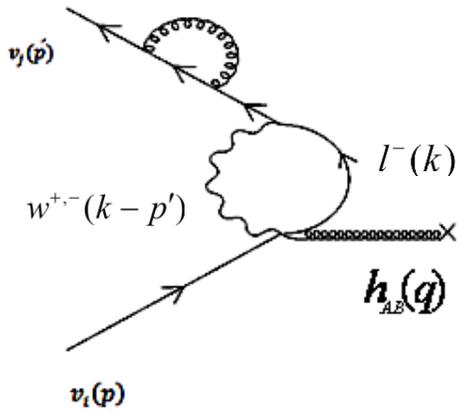

FIG. 5

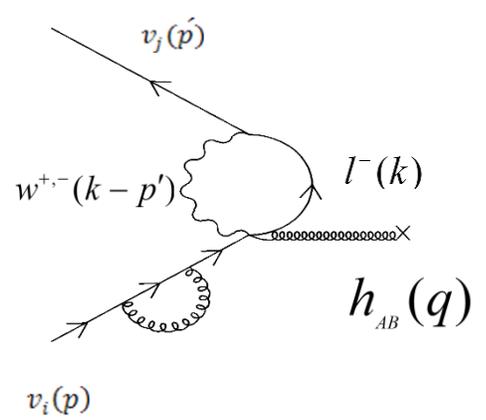

FIG. 6

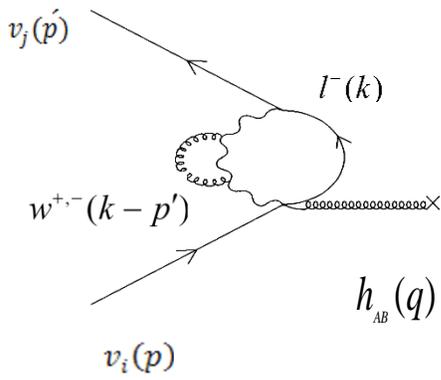

$v_j(\acute{p})$

$l^-(k)$

$w^{+,-}(k-p')$

$h_{AB}(q)$

$v_i(p)$

FIG. 7

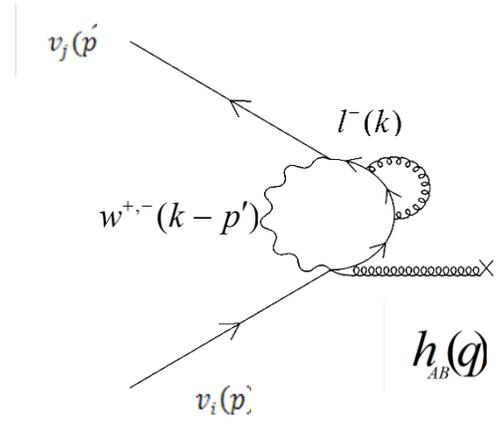

$v_j(\acute{p})$

$l^-(k)$

$w^{+,-}(k-p')$

$h_{AB}(q)$

$v_i(\acute{p})$

FIG. 8

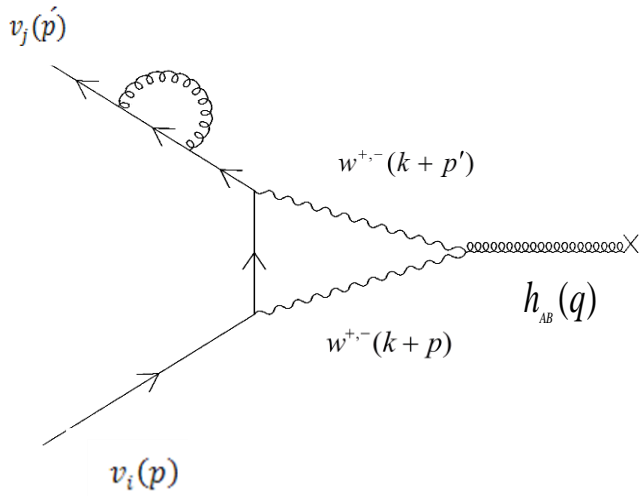

$v_j(\acute{p})$

$w^{+,-}(k+p')$

$w^{+,-}(k+p)$

$h_{AB}(q)$

$v_i(p)$

FIG. 9

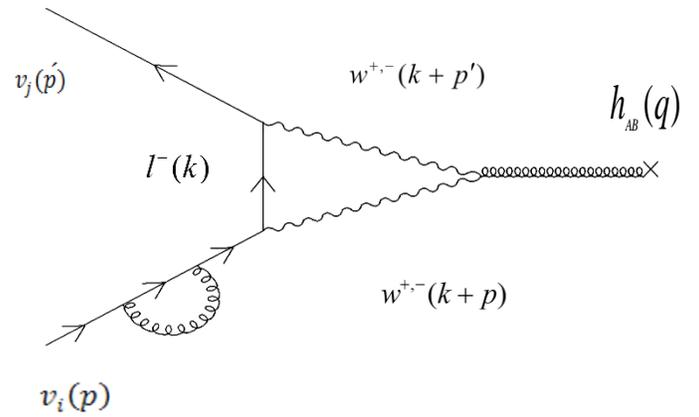

$v_j(\acute{p})$

$w^{+,-}(k+p')$

$h_{AB}(q)$

$l^-(k)$

$w^{+,-}(k+p)$

$v_i(p)$

FIG. 10

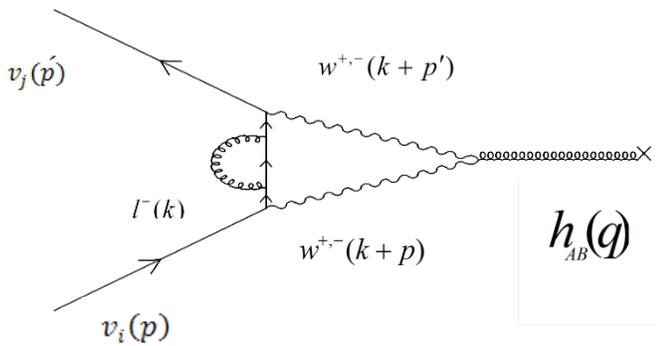

$v_j(\acute{p})$

$w^{+,-}(k+p')$

$l^-(k)$

$w^{+,-}(k+p)$

$h_{AB}(q)$

$v_i(p)$

FIG. 11

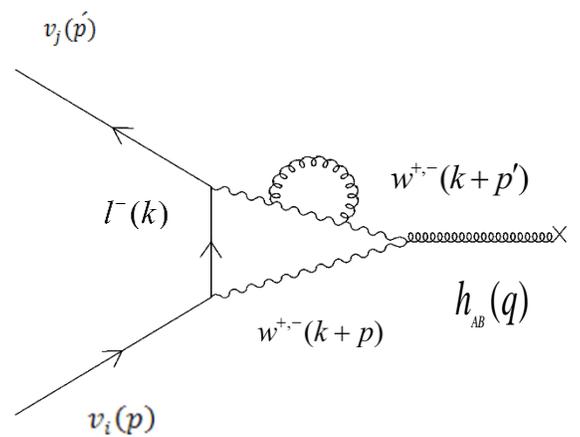

$v_j(\acute{p})$

$l^-(k)$

$w^{+,-}(k+p')$

$w^{+,-}(k+p)$

$h_{AB}(q)$

$v_i(p)$

FIG. 12

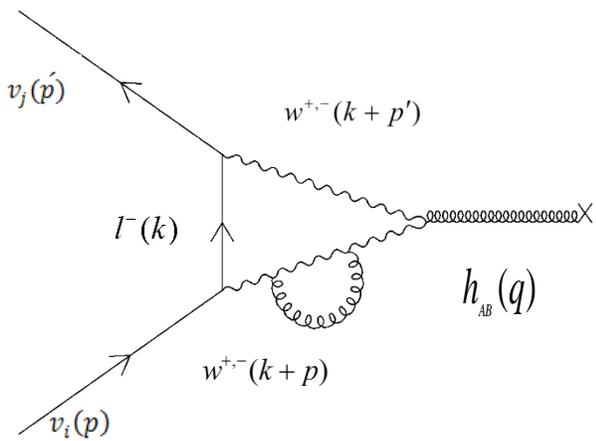

FIG. 13

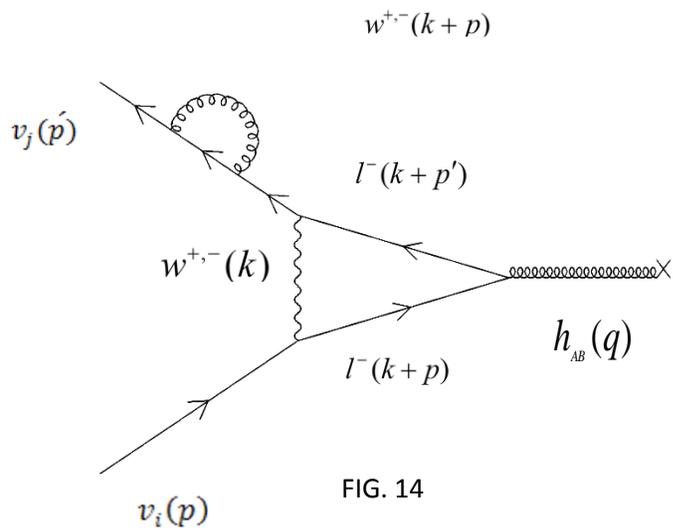

FIG. 14

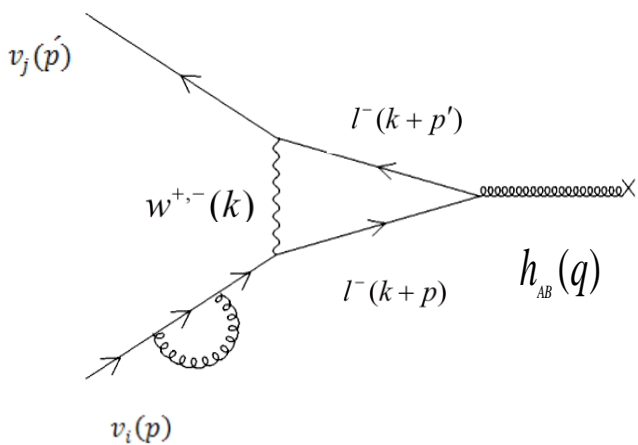

FIG. 15

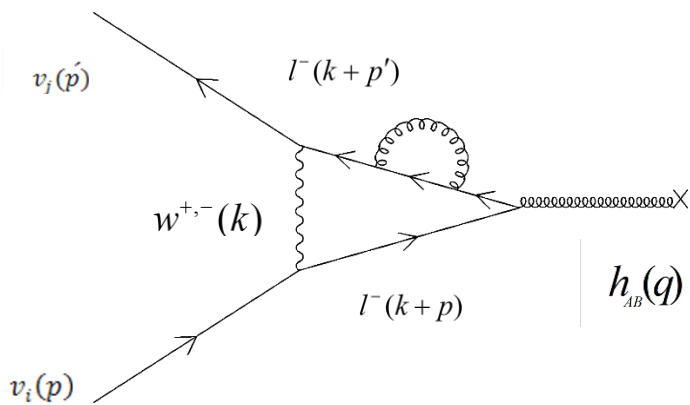

FIG. 16

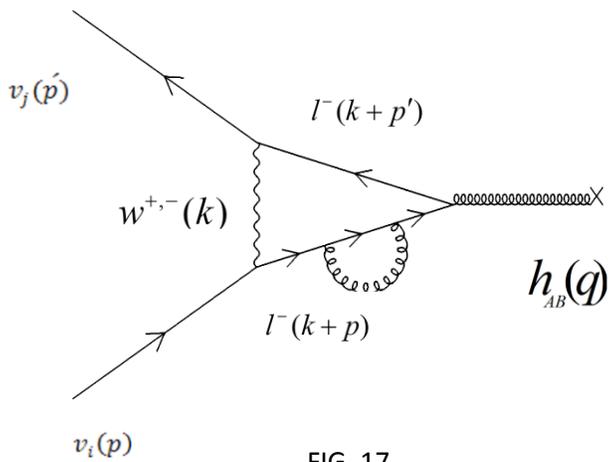

FIG. 17

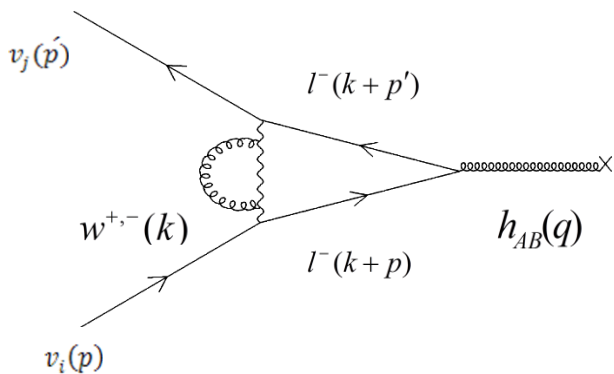

FIG. 18

The Feynman amplitudes of these diagrams are presented below. The energy-momentum four vector of the graviton propagator for the diagrams (4) and (11) is $k'$ and for the rest of the diagrams it is $p''$. The Feynman rules for the gravitational interactions with SM fields that are relevant to our study are presented in Appendix A.

Fig. 1.

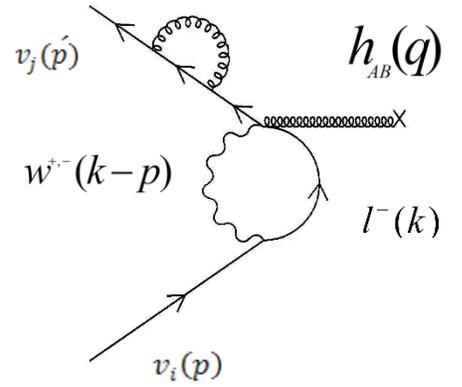

$$\bar{u}_j(p')(i\Delta J^{(1)}_{AB})u_i(p) \sim \sum_{l=e,\mu,\tau}\int\frac{d^4k}{(2\pi)^4}\int\frac{d^4p''}{(2\pi)^4}(\bar{u}_j(p'))$$

$$(-\frac{i\kappa}{8}[\gamma_{\{\mu'}(2p'-p'')_{\nu'\}}]+\frac{i\kappa}{4}\eta_{\mu'\nu'}[2p'-p''-2m_\nu])$$

$$((\eta^{\mu'\mu}\eta^{\nu'\nu}+\eta^{\mu'\nu}\eta^{\nu'\mu}-\eta^{\mu'\nu'}\eta^{\mu\nu})\times\frac{i}{2p''^2})$$

$$(i\frac{p'-p''+m_\nu}{(p'-p'')^2-m_\nu^2})(-\frac{i\kappa}{8}[\gamma_{\{\mu}(2p'-p'')_{\nu\}}]$$

$$+\frac{i\kappa}{4}\eta_{\mu\nu}[2p'-p''-2m_\nu])(i\frac{p'+m_\nu}{p'^2-m_\nu^2})(\frac{i\kappa g}{2\sqrt{2}}\Gamma_{AB\rho\beta}\gamma^\beta P_L U^*_{lj}))$$

$$(i\frac{k+m_\lambda}{k^2-m_\lambda^2})(-i\frac{\eta^{\rho\sigma}}{(k-p)^2-m_w^2})(\frac{ig}{\sqrt{2}}\gamma_\sigma P_L U_{li})(u_i(p))$$

FIG. 1

where :

$$\Gamma_{\mu\nu\rho\lambda}=[\eta_{\mu\nu}\eta_{\rho\lambda}-\frac{1}{2}(\eta_{\mu\rho}\eta_{\nu\lambda}+\eta_{\mu\lambda}\eta_{\nu\rho})]$$

and { }, denotes complete symmetrization of the indices (see Appendix C).

Fig. 2. The amplitude of this diagram can be derived from amplitude of diagram (1) by some minor changes.

Fig. 3.

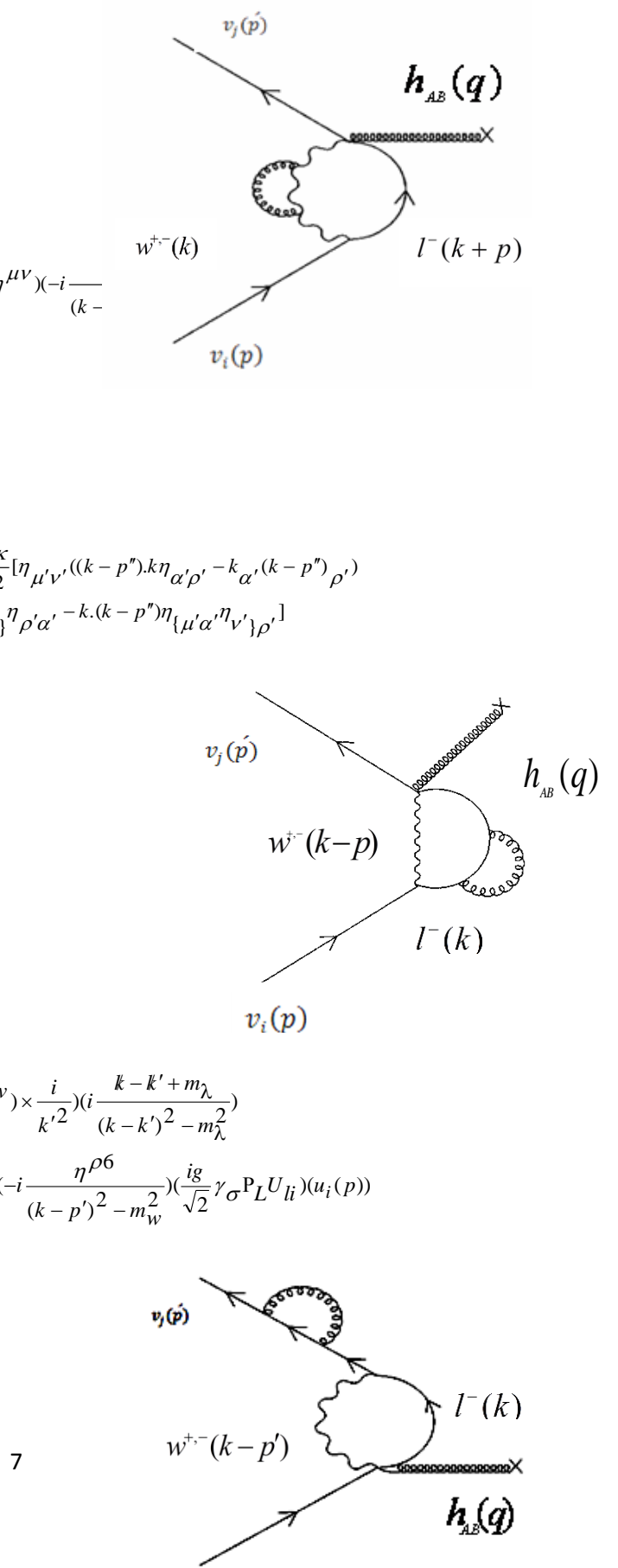

$\bar{u}_j(p')(i\Delta J_{AB}^{(3)})u_i(p) \sim \sum\limits_{l=e,\mu,\tau} \int \dfrac{d^4k}{(2\pi)^4}$

$\int \dfrac{d^4p''}{(2\pi)^4} \times (\bar{u}_j(p'))(\dfrac{i\kappa g}{2\sqrt{2}}\Gamma_{AB\beta\lambda}\gamma^\lambda P_L U^*_{lj})(\dfrac{\not{k}+\not{p}+m_l}{(k+p)^2-m_l^2})$

$(-i\dfrac{\eta^{\beta\alpha'}}{k^2-m_w^2})\times\Gamma'_{\mu'\nu'\alpha'\rho'}\times\dfrac{i}{2p''^2}(\eta^{\mu'\mu}\eta^{\nu\nu'}+\eta^{\mu'\nu}\eta^{\mu\nu'}-\eta^{\mu'\nu'}\eta^{\mu\nu})(-i-(k-$

$\times(-i\dfrac{\eta^{\varphi\rho}}{k^2-m_w^2})\times(\dfrac{ig}{\sqrt{2}}\gamma_\sigma P_L U_{li})(u_i(p))$

where :

$\Gamma'_{\mu'\nu'\alpha'\rho'}=\dfrac{i\kappa}{2}m_w^2[\eta_{\mu'\nu'}\eta_{\alpha'\rho'}-\eta_{\mu'\rho'}\eta_{\alpha'\nu'}-\eta_{\nu'\rho'}\eta_{\mu'\alpha'}]-\dfrac{i\kappa}{2}[\eta_{\mu'\nu'}((k-p'')\cdot k\eta_{\alpha'\rho'}-k_{\alpha'}(k-p'')_{\rho'})$
$+(k-p'')_{\{\mu'}k_{\alpha'}\eta_{\rho\nu'\}}+(k-p'')_{\rho'}k_{\{\nu'}\eta_{\mu'\}\alpha'}-(k-p'')_{\{\mu'}k_{\nu'\}}\eta_{\rho'\alpha'}-k\cdot(k-p'')\eta_{\{\mu'\alpha'}\eta_{\nu'\}\rho'}]$

Fig. 4.

$\bar{u}_j(p')(i\Delta J_{AB}^{(4)})u_i(p) \sim \sum\limits_{l=e,\mu,\tau} \int \dfrac{d^4k}{(2\pi)^4}\int\dfrac{d^4k'}{(2\pi)^4}(\bar{u}_j(p'))$

$(\dfrac{i\kappa g}{2\sqrt{2}}\Gamma_{AB\rho\beta}\gamma^\beta P_L U^*_{lj})(i\dfrac{\not{k}+m_l}{k^2-m_l^2})(-\dfrac{i\kappa}{8}[\gamma_{\{\mu'}(2k-k')_{\nu'\}}]+$

$\dfrac{i\kappa}{4}\eta_{\mu'\nu'}[2\not{k}-\not{k}'-2m_\nu])(\dfrac{1}{2}(\eta^{\mu'\mu}\eta^{\nu'\nu}+\eta^{\mu'\nu}\eta^{\nu'\mu}-\eta^{\mu'\nu'}\eta^{\mu\nu})\times\dfrac{i}{k'^2})(i\dfrac{\not{k}-\not{k}'+m_\lambda}{(k-k')^2-m_\lambda^2})$

$(-\dfrac{i\kappa}{8}[\gamma_{\{\mu}(2k-k')_{\nu\}}]+\dfrac{i\kappa}{4}\eta_{\mu\nu}[2\not{k}-\not{k}'-2m_\nu])(i\dfrac{\not{k}+m_\lambda}{k^2-m_\lambda^2})\ (-i\dfrac{\eta^{\rho 6}}{(k-p')^2-m_w^2})(\dfrac{ig}{\sqrt{2}}\gamma_\sigma P_L U_{li})(u_i(p))$

Fig.(5)

$$\bar{u}_j(p')(i\Delta J_{AB}^{(5)})u_i(p) \sim$$

$$\sum_{l=e,\mu,\tau} \int \frac{d^4k}{(2\pi)^4} \int \frac{d^4p''}{(2\pi)^4} (\bar{u}_j(p'))(-\frac{i\kappa}{8}[\gamma_{\{\mu'}(2p'-p'')_{\nu'\}}]$$

$$+\frac{i\kappa}{4}\eta_{\mu'\nu'}[2p'-p''-2m_\nu])$$

$$((\eta^{\mu'\mu}\eta^{\nu'\nu}+\eta^{\mu'\nu}\eta^{\nu'\mu}-\eta^{\mu'\nu'}\eta^{\mu\nu})\times\frac{i}{2p''^2})(i\frac{p'-p''+m_\nu}{(p'-p'')^2-m_\nu^2})$$

$$(-\frac{i\kappa}{8}[\gamma_{\{\mu}(2p'-p'')_{\nu\}}]+\frac{i\kappa}{4}\eta_{\mu\nu}[2p'-p''-2m_\nu])(i\frac{p'+m_\nu}{p'^2-m_\nu^2})(\frac{ig}{\sqrt{2}}\gamma_\sigma P_L U_{lj}^*)(i\frac{\not{k}+m_\lambda}{k^2-m_l^2})$$

$$(-i\frac{\eta^{\rho 6}}{(k-p')^2-m_W^2})(\frac{i\kappa g}{2\sqrt{2}}(\eta_{AB}\eta_{\rho\beta}-\frac{1}{2}\eta_{A\beta}\eta_{B\rho}-\frac{1}{2}\eta_{A\rho}\eta_{B\beta})\gamma^\beta P_L U_{li})(u_i(p))$$

Fig. 6. The amplitude of this diagram can be derived from amplitude of diagram (5) by some minor changes.

Fig. 7. The amplitude of this diagram can be derived from amplitude of diagram (3) by some minor changes.

Fig. 8. The amplitude of this diagram can be derived from amplitude of diagram (4) by some minor changes.

Fig. 9

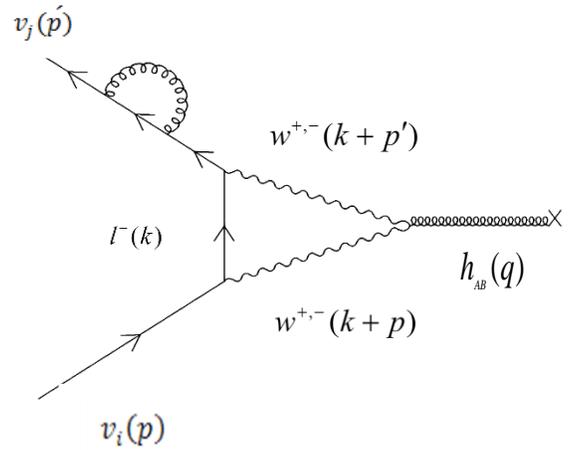

$$\bar{u}_j(p')(i\Delta J_{AB}^{(9)})u_i(p) \sim \sum_{l=e,\mu,\tau} \int \frac{d^4k}{(2\pi)^4} \int \frac{d^4p''}{(2\pi)^4} (\bar{u}_j(p'))$$

$$(-\frac{i\kappa}{8}(\gamma_{\{\mu'}(2p'-p'')_{\nu'\}}-2\eta_{\mu'\nu'}(2p'-p''-2m_\nu)))$$

$$(\eta^{\mu'\mu}\eta^{\nu'\nu}+\eta^{\mu'\nu}\eta^{\mu\nu'}-\eta^{\mu'\nu'}\eta^{\mu\nu}\frac{i}{2p''^2})$$

$$(i\frac{p'-p''+m_\nu}{(p'-p'')^2-m_\nu^2})(-\frac{i\kappa}{8}(\gamma_{\{\mu}(2p'-p'')_{\nu\}}-2\eta_{\mu\nu}(2p'-p''-2m_\nu)))$$

$$(i\frac{p'+m_\nu}{p'^2-m_\nu^2})(\frac{ig}{\sqrt{2}}\gamma_\beta P_L U_{lj}^*)(-i\frac{\eta^{\lambda\beta}}{(p'+k)^2-m_W^2})(\Gamma_{AB\rho\lambda})(-i\frac{\eta^{\rho\alpha}}{(p+k)^2-m_W^2})(i\frac{\not{k}+m_l}{k^2-m_l^2})(\frac{ig}{\sqrt{2}}\gamma_\alpha P_L U_{li})(u_i(p))$$

Fig. 10. The amplitude of this diagram can be derived from amplitude of diagram (9) by some minor changes.

Fig.11.

$$\bar{u}_j(p')(i\Delta J^{(11)}_{AB})u_i(p) \sim \sum_{l=e,\mu,\tau} \int \frac{d^4k}{(2\pi)^4} \int \frac{d^4k'}{(2\pi)^4} (\bar{u}_j(p'))(\frac{ig}{\sqrt{2}}\lambda_\beta P_L\ U^*_{lj})$$

$$(-i\frac{\eta^{\lambda\beta}}{(k+p')^2-M^2_w})(\Gamma'_{AB\rho\lambda})(-i\frac{\eta^{\alpha\rho}}{(k+p)^2-M^2_w})(i\frac{\slashed{k}+m_l}{k^2-m^2_l})$$

$$(-\frac{i\kappa}{8}([\gamma_{\{\mu}(2k-k')_{\nu\}}]-2\eta_{\mu\nu}[2\slashed{k}-\slashed{k}'-2m_l]))$$

$$(\frac{i}{2k'^2}(\eta^{\mu\mu'}\eta^{\nu\nu'}+\eta^{\mu\nu'}\eta^{\nu\mu'}-\eta^{\mu\nu}\eta^{\mu'\nu'})\times(i\frac{\slashed{k}-\slashed{k}'+m_l}{(k-k')^2-m^2_l})$$

$$(-\frac{i\kappa}{8}([(2k-k')_{\nu'}]-2\eta_{\mu'\nu'}[2\slashed{k}-\slashed{k}'-2m_l]))(i\frac{\slashed{k}+m_l}{k^2-m^2_l})(\frac{ig}{\sqrt{2}}\gamma_\alpha P_L U_{li})(u_i(p))$$

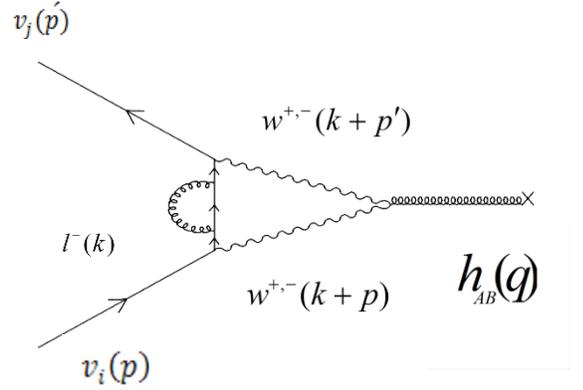

Fig.(12)

$$\bar{u}_j(p')(i\Delta J^{(12)}_{AB})u_i(p) \sim \sum_{l=e,\mu,\tau} \int \frac{dk^4}{(2\pi)^4} \int \frac{d^4p^*}{(2\pi)^4} (\bar{u}_j(p'))$$

$$(\frac{ig}{\sqrt{2}}\gamma_\beta P_L U^*_{lj})(-i\frac{\eta^{\beta\beta'}}{(k+p')^2-m^2_w})(\Gamma'_{\mu'\nu'6\beta'})$$

$$((\eta^{\mu'\mu}\eta^{\nu'\nu}+\eta^{\mu'\nu}\eta^{\mu\nu'}-\eta^{\mu'\nu'}\eta^{\mu\nu})\frac{i}{2p^{*2}})$$

$$(-i\frac{\eta^{\sigma\sigma'}}{(p'+k-p^*)^2-m^2_w})(\Gamma'_{\mu\nu\sigma'\lambda'})(-i\frac{\eta^{\lambda\lambda'}}{(p'+k)^2-m^2_w})$$

$$(\Gamma'_{AB\rho\lambda})(-i\frac{\eta^{\rho\alpha}}{(p+k)^2-m^2_w})(i\frac{\slashed{k}+m_l}{k^2-m^2_l})(\frac{ig}{\sqrt{2}}\gamma_\alpha P_L U_{li})(u_i(p))$$

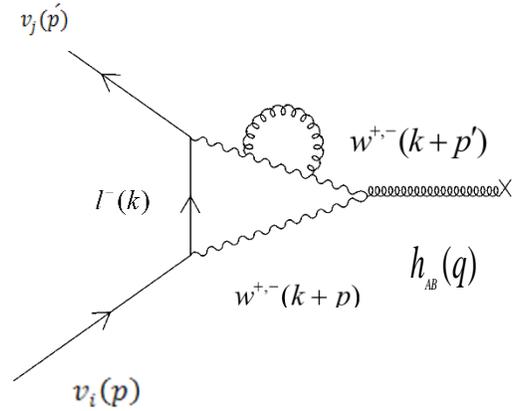

Fig. 13. The amplitude of this diagram can be derived from amplitude of diagram (12) by some minor changes.

Fig. 14.

$$\bar{u}_j(p')(i\Delta J^{(14)}_{AB})u_i(p) \sim \sum_{l=e,\mu,\tau}\sum\int\frac{d^4k}{(2\pi)^4}\int\frac{d^4p''}{(2\pi)^4}(\bar{u}_j(p'))$$

$$(-\frac{ik}{8}(\gamma_{\{\mu'}(2p'-p'')_{\nu'\}} - 2\eta_{\mu'\nu'}(2p'-p''-2m_\nu))$$

$$((\eta^{\mu'\mu}\eta^{\nu'\nu} + \eta^{\mu'\nu}\eta^{\nu'\mu} - \eta^{\mu'\nu'}\eta^{\mu\nu})\times\frac{i}{2p''^2})$$

$$(i\frac{\slashed{p}'-\slashed{p}''+m_\nu}{(p'-p'')^2-m_\nu^2})$$

$$(-\frac{ik}{8}(\gamma_{\{\mu}(2p'-p'')_{\nu\}} - 2\eta_{\mu\nu}(2p'-p''-2m_\nu)))$$

$$(i\frac{\slashed{p}'+m_\nu}{p'^2-m_\nu^2})(\frac{ig}{\sqrt{2}}\gamma_\alpha P_L U^*_{lj})(i\frac{\slashed{k}+\slashed{p}'+m_l}{(k+p')^2-m_l^2})(-\frac{ik}{8}(\gamma_{\{A}(2k+p+p')_{B\}}$$

$$-2\eta_{AB}(2k+p+p'-2m_l)))(i\frac{\slashed{k}+\slashed{p}+m_l}{(k+p)^2-m_l^2})(-i\frac{\eta^{\rho\alpha}}{k^2-m_w^2})(\frac{ig}{\sqrt{2}}\gamma_\rho P_L U_{li})(u_i(p))$$

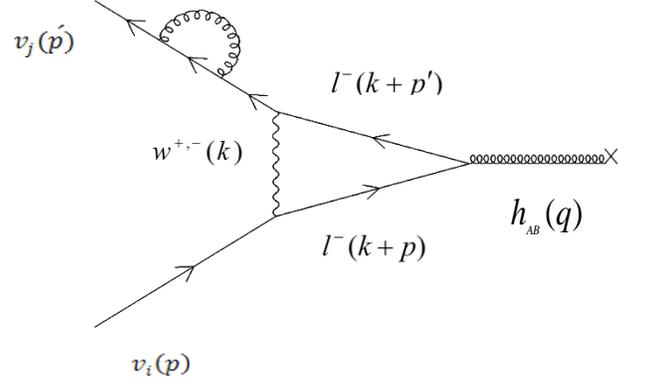

Fig. 15. The amplitude of this diagram can be derived from amplitude of diagram (14) by some minor changes.

Fig. 16.

$$\bar{u}_j(p')(i\Delta J^{(16)}_{AB})u_i(p) \sim$$

$$\sum_{l=e,\mu,\tau}\int\frac{dk^4}{(2\pi)^4}\int\frac{d^4p''}{(2\pi)^4}(\bar{u}_j(p'))(\frac{ig}{\sqrt{2}}\gamma_\alpha P_L U^*_{lj}(i\frac{\slashed{k}+\slashed{p}'+m_\lambda}{(k+p')^2-m_\lambda^2})$$

$$(-\frac{ik}{8}(\gamma_{\{\mu'}(2p'+2k-p'')_{\nu'\}} - 2\eta_{\mu'\nu'}(2p'+2\slashed{k}-p''-2m_\lambda)))$$

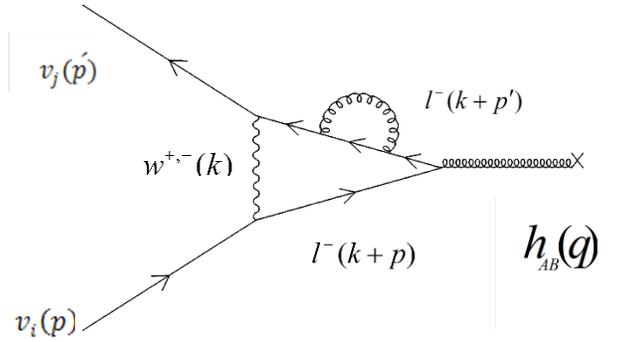

$$(\eta^{\mu'\mu}\eta^{\nu'\nu}+\eta^{\mu'\nu}\eta^{\mu\nu'}-\eta^{\mu'\nu'}\eta^{\mu\nu})(i\frac{\not{p}'+\not{k}-\not{p}''+m_\lambda}{(p'+k-p'')^2-m^2})$$

$$(\frac{-ik}{8}(\gamma_{\{\mu}(2p'+2k-p'')_{\nu\}}-2\eta_{\mu\nu}(2\not{p}'+2\not{k}-\not{p}''-2m_\lambda)))$$

$$(i\frac{\not{k}+\not{p}'+m_\lambda}{(k+p')^2-m_\lambda^2})(\frac{-i\kappa}{8}(\gamma_{\{A}(p+p'+2k)_{B\}}-2\eta_{AB}(\not{p}+\not{p}'+2\not{k}-2m_l)))$$

$$(i\frac{\not{k}+\not{p}+m_\lambda}{(k+p)^2-m_\lambda^2})(\frac{-i\eta^{\rho\alpha}}{k^2-m_w^2})(\frac{ig}{\sqrt{2}}\gamma_\rho P_L U_{li})(u_i(p))$$

Fig. 17. The amplitude of this diagram can be derived from amplitude of diagram (16) by some minor changes.

Fig.18.

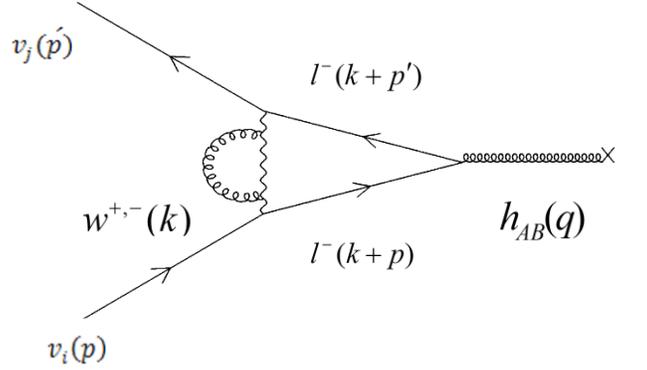

$$\bar{u}_j(p')(i\Delta J_{AB}^{(18)})u_i(p)\sim\sum_{l=e,\mu,\tau}\int\frac{dk^4}{(2\pi)^4}\int\frac{d^4p''}{(2\pi)^4}(\bar{u_j}(p'))(\frac{ig}{\sqrt{2}}\gamma_\alpha P_L U_{lj}^*)$$

$$(i\frac{\not{k}+\not{p}'+m_\lambda}{(k+p')^2-m_\lambda^2})(\frac{-i\kappa}{8}(\gamma_{\{A}(p+p'+2k)_{B\}}-2\eta_{AB}(\not{p}+\not{p}'+2\not{k}-2m_l)))(i\frac{\not{k}+\not{p}+m_\lambda}{(k+p)^2-m_\lambda^2})$$

$$(-i\frac{\eta^{\alpha\alpha'}}{k^2-m_w^2})(\Gamma'_{\alpha'\sigma'\mu'\nu'})(\eta^{\mu'\mu}\eta^{\nu'\nu}+\eta^{\mu'\nu}\eta^{\mu\nu'}-\eta^{\mu'\nu'}\eta^{\mu\nu})$$

$$(\frac{i}{2p''^2})(-i\frac{\eta^{\sigma\sigma'}}{(k-p'')^2-m_w^2})(\Gamma'_{\mu\nu\rho\sigma})(-i\frac{\eta^{\rho\rho'}}{k^2-m_w^2})(\frac{ig}{\sqrt{2}}\gamma_{\rho'}P_L U_{li})(u_i(p))$$

We have calculated the invariant Feynman amplitude for all diagrams but due to a large number of calculations we do not present the details of the calculations of all diagrams, but instead we present some calculations of two typical diagrams and write down only the final results for the rest of them. Our purpose is to find what form factors are produced by individual diagrams. Let us consider the Feynman amplitude of diagram (1). The multiplication of the numerator of the second and third parentheses regardless of the constant coefficients gives:

$$(4\gamma^\mu p'^\nu - 2\gamma^\mu p''^\nu + 4\gamma^\nu p'^\mu - 2\gamma^\nu p''^\mu - 8\eta^{\mu\nu} m_\nu + 4p'\eta^{\mu\nu} - 2p''\eta^{\mu\nu}) \tag{1}$$

and for the case of fourth and fifth parentheses we have:

$$(2p'\gamma_\mu p'_\nu - 2p''\gamma_\mu p'_\nu + 2m_\nu\gamma_\mu p'_\nu + 2p'\gamma_\nu p'_\mu - 2p''\gamma_\nu p'_\mu + 2m_\nu\gamma_\nu p'_\mu - p'\gamma_\mu p''_\nu + p''\gamma_\mu p''_\nu - m_\nu\gamma_\mu p''_\nu$$
$$- p'\gamma_\nu p''_\mu + p''\gamma_\nu p''_\mu - m_\nu\gamma_\nu p''_\mu - 4p'\eta_{\mu\nu} + 4p''\eta_{\mu\nu} p' + 2p'p''\eta_{\mu\nu} - 2p''p''\eta_{\mu\nu} + 2p''\eta_{\mu\nu} m_\nu$$
$$- 4\eta_{\mu\nu} m_\nu p'' + 4\eta_{\mu\nu} m_\nu{}^2) \tag{2}$$

By multiplying the numerators of parentheses (7), (8), (9) and (10) regardless of the constant coefficients we have :

$$= (\Gamma_{AB\rho\beta}\gamma^\beta P_L)(\not{k} + m_\lambda)(\eta^{\rho\sigma})(\gamma_\sigma P_L) \tag{3}$$
$$= (\eta_{AB}\eta_{\rho\beta} - \frac{1}{2}\eta_{A\beta}\eta_{B\rho} - \frac{1}{2}\eta_{A\rho}\eta_{B\beta})\gamma^\beta \times P_L(\not{k} + m_\lambda)(\eta^{\rho\sigma})(\gamma_\sigma P_L)$$
$$= (\eta_{AB}\gamma_\rho - \frac{1}{2}\gamma_A\eta_{B\rho} - \frac{1}{2}\eta_{A\rho}\gamma_B)(P_L)(\not{k} + m_\lambda)(\eta^{\rho\sigma})(\gamma_\sigma P_L) = (\eta_{AB}\gamma^\sigma - \frac{1}{2}\gamma_A\eta_B^\sigma - \frac{1}{2}\gamma_B\eta_A^\sigma)(\not{k})\gamma_\sigma P_L$$

But :

$$\gamma_A\not{k} = \gamma_A k^\alpha\gamma_\alpha = k^\alpha(2\eta_{A\alpha} - \gamma_\alpha\gamma_A) = 2k_A - \not{k}\gamma_A$$

So Eq. (3) takes the following form:

$$= -2\eta_{AB}\not{k}P_L - \frac{1}{2}(2k_A - \not{k}\gamma_A)\gamma_B P_L - \frac{1}{2}(2k_B - \not{k}\gamma_B)\gamma_A P_L = -2\eta_{AB}\not{k}P_L - (k_A\gamma_B + k_B\gamma_A)P_L +$$
$$\frac{1}{2}\not{k}(\gamma_A\gamma_B + \gamma_B\gamma_A)P_L \tag{4}$$
$$= -2\eta_{AB}\not{k}P_L - (k_A\gamma_B + k_B\gamma_A)P_L + \not{k}(\eta_{AB})P_L = -\eta_{AB}\not{k}P_L - (\gamma_A k_B + \gamma_B k_A)P_L$$

If we multiply parentheses (6), (7), (8), (9) and (10) we get:

$$-p'\eta_{AB}\not{k}P_L - m_\nu\eta_{AB}\not{k}P_L - p'(\gamma_A k_B + \gamma_B k_A)P_L - m_\nu(\gamma_A k_B + \gamma_B k_A)P_L \tag{5}$$

We now calculate some typical integrals appear in the calculations of diagram (1). By multiplying the first term of (1) in the first term of (2) and then multiplying the result in (5) and integration over $p''$ and $k$ we have:

$$(4\gamma^\mu p'^\nu) \times (2p'\gamma_\mu p'_\nu) = -16 p' p'^2 \qquad (6)$$

$$\int \frac{d^4k}{(2\pi)^4} \int \frac{d^4p''}{(2\pi)^4} \frac{(-16 p' p'^2) \times (-p' \slashed{k} P_L \eta_{AB})}{(p''^2)((p''-p')^2-m_\nu^2)(k^2-m_\lambda^2)((k-p)^2-m_w^2)}$$
$$= 16 p\, p'^4 \; B_0 B_1 P_L \eta_{AB} \qquad (7)$$

$$\int \frac{d^4k}{(2\pi)^4} \int \frac{d^4p''}{(2\pi)^4} \frac{(-16 p' p'^2) \times (-m_\nu \slashed{k} P_L \eta_{AB})}{(p''^2)((p''-p')^2-m_\nu^2)(k^2-m_\lambda^2)((k-p)^2-m_w^2)}$$
$$= 16 m_\nu p'^2 p' p P_L \eta_{AB} B_0 B_1 \qquad (8)$$

The coefficients $B_0, B_1$ are defined in the Appendix B. These integrals generate no form factors.

The integral

$$\int \frac{d^4k}{(2\pi)^4} \int \frac{d^4p''}{(2\pi)^4} \frac{(-16 p' p'^2) \times (-p'(\gamma_A k_B + \gamma_B k_A) P_L)}{(p''^2)((p''-p')^2-m_\nu^2)(k^2-m_\lambda^2)((k-p)^2-m_w^2)}$$
$$= 16 p'^4 \times B_0 B_1 (\gamma_A p_B + \gamma_B p_A) P_L \qquad (9)$$

and the corresponding integral of diagram (5) generate form factors $E_3$ and $D_3$. The definitions of the form factors have been presented in Appendix C.

The following integral

$$\int \frac{d^4k}{(2\pi)^4} \int \frac{d^4p''}{(2\pi)^4} \frac{(-16 p' p'^2) \times (-m_\nu(\gamma_A k_B + \gamma_B k_A) P_L)}{(p''^2)((p''-p')^2-m_\nu^2)(k^2-m_\lambda^2)((k-p)^2-m_w^2)}$$
$$= 16 p'^2 m_\nu p' \times B_0 B_1 (\gamma_A p_B + \gamma_B p_A) P_L \qquad (10)$$

also produces no form factors.

Now we study the Feynman amplitude of diagram (5) which we may call it the corresponding diagram of diagram (1). By comparing the invariant amplitude of diagrams (1) and (5) we observe that the parentheses (1) - (6) are the same in two amplitudes. By multiplying the numerators of parentheses (6), (7), (8), (9) and (10) regardless of the constant coefficient we have:

$$-p'\not{k}\ P_L\ \eta_{AB} - m_v \not{k}\ P_L\ \eta_{AB} - p'(\gamma_A\ k_B + \gamma_B\ k_A)P_L$$
$$-m_v(\gamma_A\ k_B + \gamma_B\ k_A)P_L$$

(11)

Now let us calculate some typical integrals appear in the calculations of diagram (5). By multiplying the first term of (1) in the first term of (2) and then multiplying the result in (11) and integration over $p''$ and $k$ we have:

$$(4\gamma^\mu\ p'^\nu) \times (2p'\gamma_\mu p'_\nu) = -16p'p'^2$$

(12)

$$\int \frac{d^4k}{(2\pi)^4} \int \frac{d^4p''}{(2\pi)^4}\ \frac{(-16p'p'^2) \times (-p'\not{k}\ P_L\ \eta_{AB})}{(p''^2)((p''-p')^2 - m_v^2)(k^2 - m_\lambda^2)((k-p')^2 - m_w^2)}$$
$$= 16p'\ p'^4\ B_0 B_1\ P_L\ \eta_{AB}$$

(13)

This integral produces no form factors.

$$\int \frac{d^4k}{(2\pi)^4} \int \frac{d^4p''}{(2\pi)^4}\ \frac{(-16p'\ p'^2) \times (-m_v \not{k}\ P_L\ \eta_{AB})}{(p''^2)((p''-p')^2 - m_v^2)(k^2 - m_\lambda^2)((k-p')^2 - m_w^2)}$$
$$= 16m_v\ p'^4 P_L\ \eta_{AB} B_0 B_1$$

(14)

This integral gives us $E_1$ and $D_1$ form factors.

$$\int \frac{d^4k}{(2\pi)^4} \int \frac{d^4p''}{(2\pi)^4}\ \frac{(-16p'\ p'^2) \times (-p'(\gamma_A\ k_B + \gamma_B\ k_A)P_L)}{(p''^2)((p''-p')^2 - m_v^2)(k^2 - m_\lambda^2)((k-p')^2 - m_w^2)}$$
$$= 16p'^4 \times B_0 B_1(\gamma_A\ p'_B + \gamma_B\ p'_A)\ P_L$$

(15)

This integral and the corresponding integral of diagram (1) generate form factors $E_3$ and $D_3$.

Finally the integral

$$\int \frac{d^4k}{(2\pi)^4} \int \frac{d^4p''}{(2\pi)^4}\ \frac{(-16p'\ p'^2) \times (-m_v(\gamma_A\ k_B + \gamma_B\ k_A)P_L)}{(p''^2)((p''-p')^2 - m_v^2)(k^2 - m_\lambda^2)((k-p')^2 - m_w^2)}$$
$$= 16\ p'^2 m_v\ p' \times\ B_0\ B_1(\gamma_A\ p_B + \gamma_B\ p_A)P_L$$

(16)

Produces no form factors.

We have also calculated the Feynman amplitudes of all other diagrams, which the results are as follows:

All of the diagrams generate form factors $E_3$ and $D_3$. in addition diagrams (2) ,(5) ,(9) ,(10) ,(14) and (15) generate form factors $E_1$ and $D_1$.

Interaction of neutrinos with weak gravitational fields to the first order (1-loop) involves four diagrams which produce form factors $E_3$ and $D_3$. From theoretical point of view these form factors can be used to distinguish between Dirac and Majorana neutrinos but as it is shown in [16] they satisfy in the following relations:

$$E_3^M(q^2) = E_3^D(q^2) + D_3^D(q^2) \; , \; D_3^M(q^2) = 0 \tag{17}$$

Interaction of neutrinos with gravitational fields to the second order (2-loops) involves 18 diagrams. We have shown that they produce form factors $E_1, D_1, E_3$ and $D_3$. The form factors $E_1$ and $D_1$ are new i.e. they were absent in the first order calculations, so from theoretical point of view there are more differences between Dirac and Majorana neutrinos than the first order. But corresponding to each Feynman diagram of Dirac neutrinos, there is an additional "charge conjugated" diagram for Majorana neutrinos, with $P_L \rightarrow P_R$ and $l^- \rightarrow l^+$. For electromagnetic interaction, the coupling "$e$" (electron charge) changes sign under charge conjugation but in this case (gravitational interaction),we have [16] :

$$g_{\mu\nu} \cong \eta_{\mu\nu} + k h_{\mu\nu} + O(h^2) \; , \; k = \sqrt{32\pi G} \; , \eta_{\mu\nu} = (1, -1, -1, -1)$$

where $h_{\mu\nu}$ is the spin-2 graviton, the couling $k$ does not change sign under charge conjugation, so it is the same for Majorana and Dirac neutrinos. On the other hand noting that $P_L = \frac{1}{2}(1 - \gamma_5)$ and $P_R = \frac{1}{2}(1 + \gamma_5)$, one can easily check that axial vector parts $\gamma_5$ and $-\gamma_5$ cancel each other, so in Majorana case the vertex factor do not have any terms proportional to $\gamma_5$ and therefore we do not have $D$ form factors (for more information about form factors see Appendix C). On the other hand the first terms in $P_L$ and $P_R$ do not cancel each other, so we have $E$ form factors for the Majorana neutrinos which is twice the same form factor of Dirac neutrinos. To be more clear, suppose in the Feynman amplitude we have a term proportional to $P_L$, like $AP_L$, where $A$ is a cofficient (resulting from integration, see the calculations after diagram 18). As stated before, for Majorana neutrinos there is also a term $AP_R$ in the amplitude but for Dirac case there is no term proportional to $P_R$, so mathematically the above argument can be summarized as follows :

For Majorana neutrinos : $AP_L + AP_R = \frac{1}{2}A(1 - \gamma_5) + \frac{1}{2}A(1 + \gamma_5) = \left[2 \times \frac{1}{2} + 0 \times \gamma_5\right]A.$

For Dirac neutrinos : $P_R = 0$, so $AP_L + AP_R = \left[\frac{1}{2} + \frac{1}{2}\gamma_5\right]A$.

$D_i$ and $E_i$, i =1,3 are the cofficients of the axial and non-axial parts of the amplitude respectively (see Appendic C), so we have :

$E_i^M = 2A, \; E_i^D = A \rightarrow E_i^M = 2E_i^D$

$D_i^M = 0, \quad i = 1,3$

and

$E_i^D = D_i^D = A$

So we arrive at the following relation:

$E_i^M = E_i^D + D_i^D, \qquad i = 1,3$ (19)

This is an important result and show that despite theoretical differences we are not able to distinguish experimentally Majorana neutrinos from Dirac neutrinos interacting with gravity. Let us explain it more clearly by a simple example. Suppose we get the value 4 in an experiment for the matrix element $\bar{u}_j(p')(i\Delta J_{AB})u_i(p)$. But we know 4=2+2, so experimentally one cannot distinguish whether we have obtained 4 (the left-hand side) or 2+2 (the right-hand side). 4 and (2+2) correspond to Majorana and Dirac neutrinos respectively. Therefore in spite of theoretical differences in the graviton vertex of the two cases, one would not be able to distinguish Majorana and dirac neutrinos experimentally.

## Discussion

Equivalence principle (EP) is one of the cornerstones of general relativity. Considerable efforts have been made and are still being made to test the EP for antimatter.

There are some direct experiments and observations which indicate EP holds also for antimatter and the interactions between matter and antimatter are the same as those between matter and itself, so matter and antimatter behave identically in the gravitational fields.

The famous worldwide experiments to test the equivalence principle for antimatter (independent of its composition or structure) with very high precision are as follows:

1). ALPHA- Antihydrogen Laser Physics Apparatus[30].

2). AEGIS- Antihydrogen Experiment : Gravity, Interferometry, Spectroscopy [31].

3). GBAR- Gravitational Behavior of Antihydrogen at Rest [32].

All these three facilities rely on the Antiproton Decelerator (AD) at CERN, but AEGIS and GBAR use beams of antihydrogen rather than trapped antihydrogen.

4). AGE- Antimatter Gravity Experiment at Fermilab [33].

They made measurements directly testing both the equivalence principle and that matter and antimatter behave identically in the gravitational field of the earth.
Using the gravitationally coupled Dirac equation it is shown in [34] that particles and antiparticles experience the same coupling to the gravitational field, including all relativistic quantum corrections of motion. Their investigations demonstrate the consistency of quantum mechanics with general relativity and suggest that any conceivable differences of the gravitational coupling of particles and antiparticles should be assigned to a "fifth force," not to any conceivable "modifications of the gravitational mass" of antiparticles versus particles.
On the other hand, in particular there is an observational confirmation for neutrinos and antineutrinos. Based on data from the supernova SN 1987A it is confirmed that the Einstein equivalence principle is valid for electron neutrinos and their antiparticles [35].
So the experimental and observational data show that gravity cannot distinguish between matter and antimatter. We know that for Majorana neutrinos, the neutrinos and antineutrinos are the same but for Dirac case, they are different. If gravity cannot distinguish matter from antimatter it cannot distinguish Majorana and Dirac neutrinos.

## Conclusions

We have studied the graviton-neutrino vertex and gravitational neutrinos transition form factors to the second order i.e. at 2-loops. We have shown that from theoretical point of view the second order calculations reveal more differences between Dirac and Majorana neutrinos than the first order but we have also shown that they are indistinguishable as long as experiments are considered. As it is well known the most sensitive way to distinguish Majorana from Dirac neutrinos is neutrinoless double beta decay which is too far from the realm of gravity.

**Acknowledgement.**

S. A. Alavi would like to thank the Department of Physics of the University of Torino (Italy) for hospitality where some parts of this work were done. He also very grateful to Samoil Bilenkey (JINR Dubna) and Carlo Giunti (INFN, Torino) for useful discussions.

**References**

[1]. S. A. Alavi, S. F. Hosseini, Gravitation and Cosmology **19**, 129 (2013); arXiv:1108.3593.

[2]. S. A. Alavi, S. Nodeh, Phys. Scr. **90**, 035301 (2015); arXiv:1301.5977.

[3]. K. Kohkichi, M. Kasai, Prog. Theor. Phys. **100**, 1145 (1998); arXiv :0603035.

[4]. Y.N. Obukhov, A. J. Silenko, Oleg. V. Teryaev, Phys.Rev.D **84**, 024025 (2011); arXiv:1106.0173 and Phys. Rev. D **88**, 084014 (2013); arXiv:1308.4552.

[5]. B. Mashhoon, Lect. Notes Phys.**702,** 112 (2006); arXiv:hep-th/0507157.

[6]. G. Lambiase, G. Papini, R. Punzi and G. Scarpetta, Phys.Rev.D **71**, 073011 (2005); arXiv:gr-qc/0503027.

[7]. R. M. Crocker, Carlo Giunti, and Daniel J. Mortlock , Phys.Rev. D **69**, 063008 (2004); arXiv:hep-ph/0308168.

[8]. Y. N. Obukhov, Phys.Rev.Lett. **86**, 192 (2001); arXiv:gr-qc/0012102, Fortsch.Phys. **50**, 711 (2002); arXiv:gr-qc/0112080.

[9]. G. Z. Adunas, E. Rodriguez-Milla, D. V. Ahluwalia, Gen.Rel.Grav.**33,** 183 (2001); arXiv:gr-qc/0006022 .

[10]. Michael J. W. Hall, Gen.Rel.Grav.**37,** 1505 (2005); arXiv:gr-qc/0408098.

[11]. Hans-Juergen Matschull, Max Welling, Class.Quant.Grav. **15,** 2981 (1998); arXiv:gr-qc/9708054.

[12]. Arkady L. Kholodenko, Adv.Stud.Theor.Phys. **4,** 689 (2010); arXiv:1003.5696.

[13]. Abhay Ashtekar, Annalen Phys. **9,** 178 (2000);  arXiv:gr-qc/9910101.

[14]. P. Hernandez, CERN Yellow Report CERN **001**, 229 (2010).

[15]. N. E. Mavromatos, CERN-PH-TH/**252** (2011), KCL-PH-TH/**34** (2011), LCTS/**17** (2011); arXiv : 1110.3729.

[16]. A. Menon, Arun M. Thalapillil, Phys.Rev.D **78**, 113003 (2008).

[17]. K. L. Ng,  Phys.Rev. D **47,** 5187 (1993).

[18].  José F. Nieves, Palash B. Pal, Phys.Rev.D **77,** 113001 (2008).

[19]. D. Singh, et.al; Phys. Rev. Lett. **97**, 041101 (2006).

[20]. José F. Nieves, Palash B. Pal, Phys. Rev. Lett. **98,** 069001 (2007).

[21]. D. Singh, et.al; Phys. Rev. Lett. **98,** 069002 (2007).

[22]. Zhi-zhong Xing,  He Zhang,  Commun.Theor.Phys. **48**, 525 (2007).

[23]. Thomas D. Gutierrez , Phys. Rev. Lett. **96**, 121802 (2006).

 [24].  M. Zralek,  Acta Phys.Polon.B **28**, 2225 (1997).

[25].  V.B. Semikoz, : Nucl.Phys. B **498**, 39 (1997).

[26].  Steen H. Hansen,  Report  Number : TAC-**30** (1997); arXiv:hep-ph/9708359.

[27].  G. V. Dass, Phys. Rev. D. **32,** 1239 (1985).

[28].  B. Kayser, Phys. Lett. B **112**, 137 (1982).

[29].  J. Barranco et.al., arXiv:1408.3219, DOI : 10.1016/j.physletb.2014.11.008.

[30]. The  ALPHA Collaboration & A.E. Charman, Nature  Communications **4**, 1 (2013).

[31]. AEgIS Collaboration, JINST **8**, 08013 (2013); arXiv :1306.5602.

[32]. P. Debu, Journal of Physics: Conference Series **460,** 012008 (2013).

[33]. Z. Yaofu, American Physical Society, APS April Meeting  2010, February 13-16, 2010, abstract

   #R1.018, Bibliographic Code : 2010APS.APR.R1018Z.

[34]. U. D. Jentschura, Phys. Rev. A **87,** 032101 (2013).

[35]. S. Pakvasa, W. A. Simmons and T. J. Weiler, Phy. Rev. D **39**, 1761 (1989).

## Appendix A.

Some Feynman rules for gravitational interactions with SM fields that are relevant to the processes considered in this article.

The Feynman rule of graviton propagator is given by :

$$P^{\mu\nu\alpha\beta}(q^2) = \frac{1}{2q^2}\left(\eta^{\mu\alpha}\eta^{\nu\beta} + \eta^{\nu\alpha}\eta^{\mu\beta} - \eta^{\mu\nu}\eta^{\alpha\beta}\right)$$

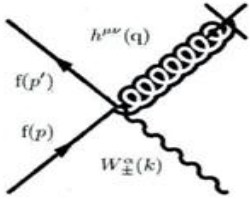

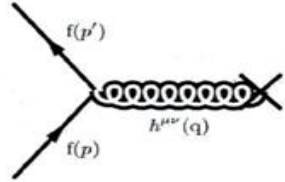

Graviton-fermion coupling is $\frac{-i\kappa}{8}[\gamma_{\{\mu}(p+p')_{\nu\}}] + \frac{i\kappa}{8}\eta_{\mu\nu}[\not{p}+\not{p}'-2m_f]$

For the four point graviton coupling the Feynman rule is $i\frac{\kappa g}{2\sqrt{2}}[\eta_{\mu\nu}\eta_{\alpha\beta} - \frac{1}{2}\eta_{\mu\beta}\eta_{\nu\alpha} - \frac{1}{2}\eta_{\mu\alpha}\eta_{\nu\beta}]\gamma^\beta P_L$

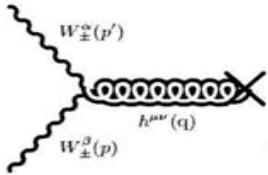

The Feynman rule is $i\kappa\frac{M_W^2}{2}[\eta_{\mu\nu}\eta_{\alpha\beta} - \eta_{\mu\alpha}\eta_{\nu\beta} - \eta_{\nu\alpha}\eta_{\mu\beta}] - i\frac{\kappa}{2}[\eta_{\mu\nu}(p'\cdot p - p_\alpha p'_\beta) + p'_{\{\mu}p_\alpha\eta_{\beta\nu\}} + p'_\beta p_{\{\nu}\eta_{\mu\}\alpha} - p'_{\{\mu}p_{\nu\}}\eta_{\alpha\beta} - (p'\cdot p)\eta_{\{\mu\alpha}\eta_{\nu\}\beta}]$

# Appendix B.

The definition of $B_0$ is as follows :

$$B_0 = \int \frac{1}{p''^2((p''-p')^2-m_v^2)} d^4p'' = \int_0^1 dx \int \frac{1}{(p''^2-\alpha)^2} d^4p'' = \int_0^1 dx \; I_2(\alpha)$$

where :

$$\alpha = x^2 p'^2 - x(p'^2 - m_v^2)$$

$$I_n(\alpha) = \int d^D q \frac{1}{(q^2-\alpha)^n} = i(-1)^n \pi^{D/2} \frac{\Gamma(n-D/2)}{\Gamma(n)} \alpha^{D/2-n-\varepsilon}$$

Now we define $B^\rho$ as :

$$B^\rho = \int \frac{k^\rho}{(k^2-m_\lambda^2)((k-p)^2-m_w^2)} d^4k$$

But :

$$\frac{1}{ab} = \int_0^1 dx \frac{1}{(a(1-x)+bx)^2}$$

$$a = k^2 - m_\lambda^2 \;\; , \;\; b = (k-p)^2 - m_w^2$$

$$\Rightarrow B^\rho = \int_0^1 dx \int_k k^\rho [(k^2-m_\lambda^2)(1-x)+((k-p)^2-m_w^2)x]^{-2}$$

By introducing the new variables $A$ and $k'$ as :

$$A = x^2 p^2 - x(p^2 - m_w^2 + m_\lambda^2) + m_\lambda^2$$

$$k' = k - x\,p$$

$B^\rho$ can be rewritten as follows :

$$B^{\rho} = \int_0^1 dx \int (k' + xp)^{\rho} (k'^2 - A)^{-2} d^4 k'$$

So :

$$B^{\rho} = \int_0^1 dx \int \frac{(k' + xp)^{\rho}}{(k'^2 - A)^2} d^4 k' = \int_0^1 dx \int \frac{x\, p^{\rho}}{(k'^2 - A)^2} d^4 k'$$

Therefore we have :

$$B^{\rho} = p^{\rho} \times B_1$$

where :

$$B_1 = \int_0^1 dx \int \frac{x}{(k'^2 - A)^2} d^4 k' = \int_0^1 dx\, x\, I_2(A)$$

## Appendix C.

In this part we present the definitions of the form factors. We denote by $\Gamma_{\mu\nu}$ (p, q) the gravitational vertex function of neutrino which is symmetric in its indices. The gravitational gauge invariance implies that it satisfies the condition :

$$q^{\mu} \bar{u}(p') \Gamma_{\mu\nu} u(p) = 0$$

In order to write down the general form of the matrix element, we introduce its tensor and pseudotensor components as follows :

$$\bar{u}(p') \Gamma_{\mu\nu} u(p) = \bar{u}(p') \big[ \Gamma'_{\mu\nu} + \Gamma''_{\mu\nu} \gamma^5 \big] u(p) \tag{C.1}$$

Lorentz invariance implies that the vertex function in general can have the following components:

$$g_{\mu\nu}\ ,\, p_{\mu} p_{\nu}\ ,\, q_{\mu} q_{\nu}\ ,\, \{pq\}_{\mu\nu}\ ,\, \{p\gamma\}_{\mu\nu}\ ,\, \{q\gamma\}_{\mu\nu}$$

where $q = p - p'$ and $\{pq\}_{\mu\nu} \equiv (p_{\{\mu} q_{\nu\}}) = p_{\mu} q_{\nu} + q_{\mu} p_{\nu}$. Therefore the tensor and pseudotensor (parity conserving and parity violating) components of neutrino gravitational vertex in general have the following forms :

$$\bar{u}(p')_j \Gamma'_{\mu\nu} u(p)_i \cong \bar{u}(p')_j [E_1(q^2)(q^2 g_{\mu\nu} - q_\mu q_\nu) + E_2(q^2)(p_\mu p_\nu) +$$

$$E_3(q^2)(q^2\{\gamma p\}_{\mu\nu} - \Delta_{ij} m_\nu \{pq\}_{\mu\nu})] u(p)_i + O(\Delta_{ij} m_\nu{}^2) \qquad \text{(C.2)}$$

$$\bar{u}(p')_j \Gamma''_{\mu\nu} \gamma^5 u(p)_i \cong \bar{u}(p')_j [D_1(q^2)\gamma^5(q^2 g_{\mu\nu} - q_\mu q_\nu) + D_2(q^2)\gamma^5(p_\mu p_\nu) +$$

$$D_3(q^2)\gamma^5(q^2\{\gamma p\}_{\mu\nu} - \Sigma_{ij} m_\nu \{pq\}_{\mu\nu})] u(p)_i + O(\Delta_{ij} m_\nu{}^2) \qquad \text{(C.3)}$$

where :

$$\Delta_{ij} m_\nu = m_{\nu j} - m_{\nu i}$$
$$\sum_{ij} m_\nu = m_{\nu j} + m_{\nu i}$$
$$\Delta_{ij} m_\nu^2 = m_{\nu j}^2 - m_{\nu i}^2$$

According to Eq.(C.2), the coefficient of $(q^2 g_{\mu\nu} - q_\mu q_\nu)$, $(p_\mu p_\nu)$ and $(q^2\{\gamma p\}_{\mu\nu} - \Delta_{ij} m_\nu\{pq\}_{\mu\nu})$, are $E_1(q^2)$, $E_2(q^2)$ and $E_3(q^2)$ respectively.

It is also seen from Eq.(C.3) that the coefficient of $\gamma^5(q^2 g_{\mu\nu} - q_\mu q_\nu)$, $(p_\mu p_\nu)\gamma^5$ and $\gamma^5(q^2\{\gamma p\}_{\mu\nu} - \Sigma_{ij} m_\nu\{pq\}_{\mu\nu})$, are $D_1(q^2)$, $D_2(q^2)$ and $D_3(q^2)$ respectively.